\documentstyle[12pt,epsf,psfig]{article}
\begin{document}
\title{\bf  The anomalous lepton magnetic moment, LFV decays  and  the fourth
generation}
\author{ Wujun Huo
 \\
{\small\it  Department of  Physics, Peking University, Beijing
$100871$,
P.R. China}\\
Tai-Fu Feng \\
{\small\it  Institute of Higy Energy Physics, CAS, P.O. Box
918(4), Beijing 100039, P.R.China}}
\date{}
\maketitle

\begin{abstract}
We investigate the lepton flavor violation (LFV) decays, $\tau\to
l\gamma$ ($l=\mu, e$) and $\mu\to e\gamma$, and the newly observed
muon $g-2$ anomaly in the framwork of a squential fourth
generation model with a heavy fourth neutrino, $\nu'$. By using
the recent experimental bounds, we take the constraints of the
$4\times 4$ leptonic mixing matrix element factors, $|V_{1\nu'}
V_{2\nu'}|^2$, $|V_{1\nu'} V_{3\nu'}|^2 $ and $|V_{3\nu'}
V_{2\nu'}|^2$. We find that LFV decays and $g_\mu -2$ can exclude
most of the parameter space of the 4th generation neutrino mass
$m_{\nu'}$ and give stringent constraints on the existence of the
fourth generation.

\end{abstract}
\newpage

At present, Standard model (SM) has to face  the experimental
difficaties which are all relate to leptons. It seems to indicate
the presence of new physics just round the cornor will be in the
leptonic part. Firstly, there are  convincing evdences that
neutrinos are massive and oscillate in flavor \cite{neutrino}.
Secondly, the recent measurment of the muon anomalous magnetic
moment by the experiment E821 \cite{e821} at Brookhaven National
Laboratory disagrees with the SM expectations at more than
2.6$\sigma$  level (the new deviation drops to  to
1.6$\sigma$\cite{new}).

Defined as $a_\mu \equiv (g_\mu -2)/2$, the recent measurement of $a_\mu$ is
\begin{eqnarray}
a^{\rm exp}_\mu =(11\,659\,202\,\pm14\pm6)\times 10^{-10},
\end{eqnarray}
where the SM prediction is
\begin{eqnarray}
a^{\rm SM}_\mu =(11\,659\,159.7\,\pm6.7)\times 10^{-10},
\end{eqnarray}
Thus, one finds,
\begin{eqnarray}
\delta a^{\rm SM}_\mu \equiv a^{\rm exp}_\mu -a^{\rm
SM}_\mu =(42.6\pm16.5)\times 10^{-10}
\end{eqnarray}
which gives the old $2.6\sigma$ deviation. But the revised
difference between experiment and SM is
 \begin{eqnarray}
\delta a^{\rm SM}_\mu \equiv a^{\rm exp}_\mu -a^{\rm SM}_\mu
=26(16)\times 10^{-10}
\end{eqnarray}
which is now only a $1.6\sigma$ deviation. Although the deviation
drops from $2.6\sigma$ to $1.6\sigma$, it might seem that the
absolute magnitude of the deviation may be a hint of new physics.
There have been a lot of scenarios of new physics proposed to
interpret the non-vanishing and positive value of $\delta a_\mu$.
Many of these new physics effects can also contribute to the
lepton flavor violation processes, such as $\tau\to l\gamma$
($l=\mu, e$) and $\mu\to e\gamma$. In this note, we consider the
sequential fourth generation standard model (SM4) to investigate
its contributions to $a_\mu$  and some leptonic flavor violation
decays.

   From the point of phenomenology,
 there is a realistic question what are numbers of the fermions
generation or weather there are other additional quarks or leptons. The
present experiments  tell us there are only three generation fermions with
$light$ neutrinos which mass are less smaller than $M_Z /2$\cite{Mark}. But
the experiments don't  exclude the existence of other additional generation,
such as the fourth generation, with a $heavy$ neutrino, i.e. $m_{\nu_4} \geq
M_Z /2$\cite{Berez}. Many refs. have studied  the SM4 \cite{McKay}, which is
added   an up-like quark $t^{'}$, a down-like quark $b^{'}$, a lepton
$\tau^{'}$,    and a heavy neutrino $\nu^{'}$ in the SM. The properties of
these new    fermions are all the same as their corresponding counterparts
  of other three generations except their masses and CKM mixing, see Tab.1,
\begin{table}[htb]
\begin{center}
\begin{tabular}{|c||c|c|c|c|c|c|c|c|}
\hline
& up-like quark & down-like quark & charged lepton &neutral lepton \\
\hline
\hline
& $u$ & $d$& $e$ & $\nu_{e}$ \\
SM fermions& $c$&$s$&$\mu$&$\nu_{\mu}$ \\
& $t$&$b$&$\tau$&$\nu_{\tau}$\\
\hline
\hline new
fermions& $t^{'}$&$b^{'}$&$\tau^{'}$&$\nu'$ \\
\hline
\end{tabular}
\end{center}
\caption{The elementary particle spectrum of SM4}
\end{table}

If there exists a very heavy fourth neurino $\nu'$,  it can
contribute to $a_\mu$ and lepton flavor violation decays through
diagram of Fig. 1. This is the electroweak interaction. Similar to
that of quarks, the corresponding Lagragian is
\begin{equation}
{\cal L}=-\frac{g}{\sqrt{2}}({\bar \nu'}\gamma_\mu a_L V_{l\nu'}
l) W^\mu +h.c.
\end{equation}
where $a_L =(1-\gamma_5)/2$, $V_{l\nu'} (l=e\mu,\tau)$ is the
$(1,4),(2,4)$ and $(3,4)$ elements of the four-generaton lepton
mixing MNS matrix ($4\times 4$),
 \begin{equation}
V^{\rm SM4}_{\rm MNS} = \left (
\begin{array}{lcrr}
V_{1\nu_e} & V_{1\nu_\mu} & V_{1\nu_\tau} & V_{1\nu'}\\
V_{2\nu_e} & V_{2\nu_\mu} & V_{2\nu_\tau} & { V_{2\nu'}}\\
V_{3\nu_e} & V_{3\nu_\mu} & V_{3\nu_\tau} & V_{3\nu'}\\
V_{4\nu_e} & V_{4\nu_\mu} & V_{4\nu_\tau} & V_{4\nu'}\\
\end{array} \right )
\end{equation}

 Reverting back to the diagrams of Fig. 1, we see that the
fourth neutrino contribution to decays $\tau\to e\gamma,\mu\gamma$
and $\mu\to e\gamma$ (see Fig. 1). We don't consider the three
usual neutrinos' contribution because their masses are too small
mass and have almost no effects on LFV decays and anomalous
magnetic moment of leptons. Because of its heavy mass, the 4th
neutrino $\nu'$ in SM4 can induce these decays. Calculating Fig.
1, we obtain
\begin{equation}
{\rm \Gamma}(\tau\to\mu\gamma) =\frac{\alpha G_F^2 m^5_\tau}{96}
|V_{2\nu'} V_{3\nu'}|^2 f^2  (x). \label{taumu}
\end{equation}
where $x\equiv m^2_{\nu'} /m^2_W$, $\alpha$ is the fine
constrcture constant and
\begin{eqnarray}
 f(x)=\frac{-5x^3 -5x^2 +4x}{12(x-1)^3}+\frac{(2x^3 -x^2 )\log x}{2(x-1)^4}.
\end{eqnarray}
Simlarly, we obtain
\begin{equation}
{\rm \Gamma}(\tau\to e\gamma) =\frac{\alpha G_F^2 m^5_\tau}{96}
|V_{1\nu'} V_{3\nu'}|^2 f^2  (x), \label{taue}
\end{equation}
\begin{equation}
{\rm \Gamma}(\mu\to e\gamma) =\frac{\alpha G_F^2 m^5_\mu}{96}
|V_{2\nu'} V_{1\nu'}|^2 f^2  (x). \label{mu}
\end{equation}
Using the current experimental bounds of these three LFV processes
\cite{pdg}, we obtain the parameter space of other $4\times 4$
matrix elements of leptonic mixing, (see Fig. 2). From  Figs. 2,
considering the unitarity of the matrix $V^{SM4}_{\rm MNS}$, we
find the reasonable  range of $m_\mu'$ is under the curve.

The  Lagragian related to $g_\mu -2$  is
\begin{equation}
{\cal L}=-\frac{g}{\sqrt{2}}({\bar \nu'}\gamma_\mu a_L V_{2\nu'}
\mu) W^\mu +h.c.
\end{equation}
From this Lagragian, we see that the fourth neutrino contribution
to $a_\mu$ is
\begin{eqnarray}
a^{\rm SM4}_\mu &=& \alpha(\frac{g}{\sqrt{2}}V_{2\nu'})^2
\frac{2m^2_\mu}{m^2_W}\cdot f(x).
\end{eqnarray}

We suppose that the 1.6$\sigma$ discrepancy of muon anomalous
magnetic moment, $\delta a^{\rm SM}_\mu$, is induced by the fourth
sequential neutrino $\nu'$. We can  use the above equation to get
parameter space of $f(x)$ and $V_{2\nu'}$ to $m_{\nu'}$ (see Fig.
3 and 4),
\begin{eqnarray}
(V_{2\nu'})^2 =\frac{{\sqrt{2}}\cdot \delta a^{\rm SM}_\mu}{
8G_{\rm F} \alpha m^2_\mu\cdot f(x)}
 \end{eqnarray}

From Fig. 3, we can see that  $f(x)$ is taken negative values
except for a very narrow range of $m_{\nu'}$ which is  from 58 GeV
to 80 GeV. In other words, the sign of $a^{\rm SM4}_\mu$ is only
related to $\nu$ mass.  Only in this narrow mass range, $\nu'$
gives positive contribution to $g_\mu -2$. The low bound of
$m_{\nu'}$ we get from $g_\mu -2$ is consistent with the present
experiments \cite{Mark}. Let's look at the parameter space of the
MNS matrix element factors again. Considering the constraint from
$g_\mu -2$, we see that the upper bounds of $m_{\nu'}$ don't
exceed 80 GeV. Although the low bound of $m_{\nu'}$ we get from
$l\to l'\gamma $ is consistent with the present experiments
\cite{Mark}, the upper bound of $m_{\nu'}$, $m_{\nu'}< 80 {\rm
GeV}$, seems to conflict with the current experiments statue which
there is no any new physics signals upper to several GeVs. The
fourth generation particles seems not to be so light.  They should
be several hundred GeVs weight.Moreover, from Fig. 4,  if we
consider the unitarity of the matrix $V^{SM4}_{\rm MNS}$ and tiny
values of its elements, the reasonable value range of $m_\nu'$
will be more narrow.

 In summary, we calculate the contribution of the fourth generation
to $g-2$ and LFV decays and get an interesting result which we can
exclude most values of $m_{\nu'}$.  Considering the current
experimental bounds,
 we give the parameter space of $m_{\nu'}$ and lepton mixing
matrix element $V_{l\nu'}$. We find that LFV processes can
constrain on the neutrino mass of the fourth generation: i.e. its
mass should be lighter than 80 GeV. It seems that from the lepton
part, the current experiments can impose a stringent constraint on
the existence of the fourth generation.

\section*{Acknowledgments}
This research is supported  by the the Chinese Postdoctoral
Science Foundation and CAS K.C. Wong Postdoctoral Research Award  Fund.
 We are grateful to prof. C.S. Huang and Prof. X.M
Zhang for useful discussions.

\newpage
\begin{figure}
\vskip -4cm \epsfxsize=20cm \epsfysize=20cm \centerline{
\epsffile{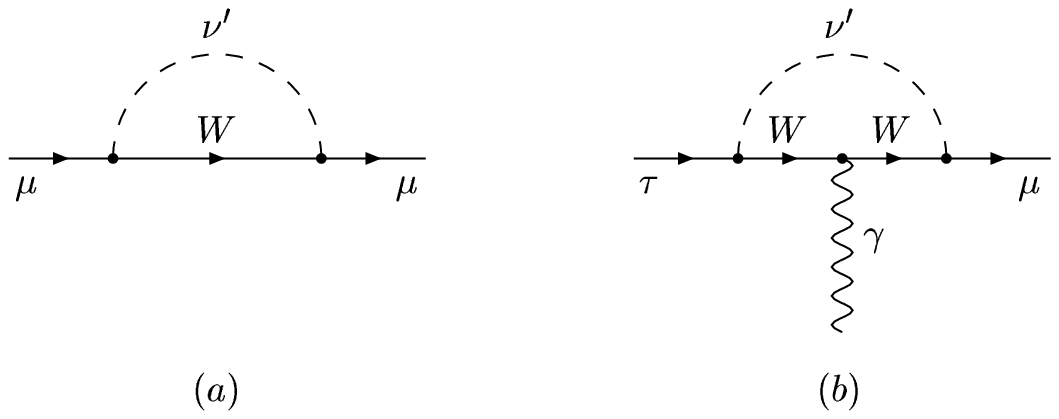}}\vskip -12cm
 \caption{Feynmann diagrams for (a) $\nu$
anomalous  magnetic moment and (b) LFV decays  with  the fourth
generation. }
\end{figure}

\begin{figure}
\vskip 1cm \epsfxsize=20cm \epsfysize=10cm \centerline{
\epsffile{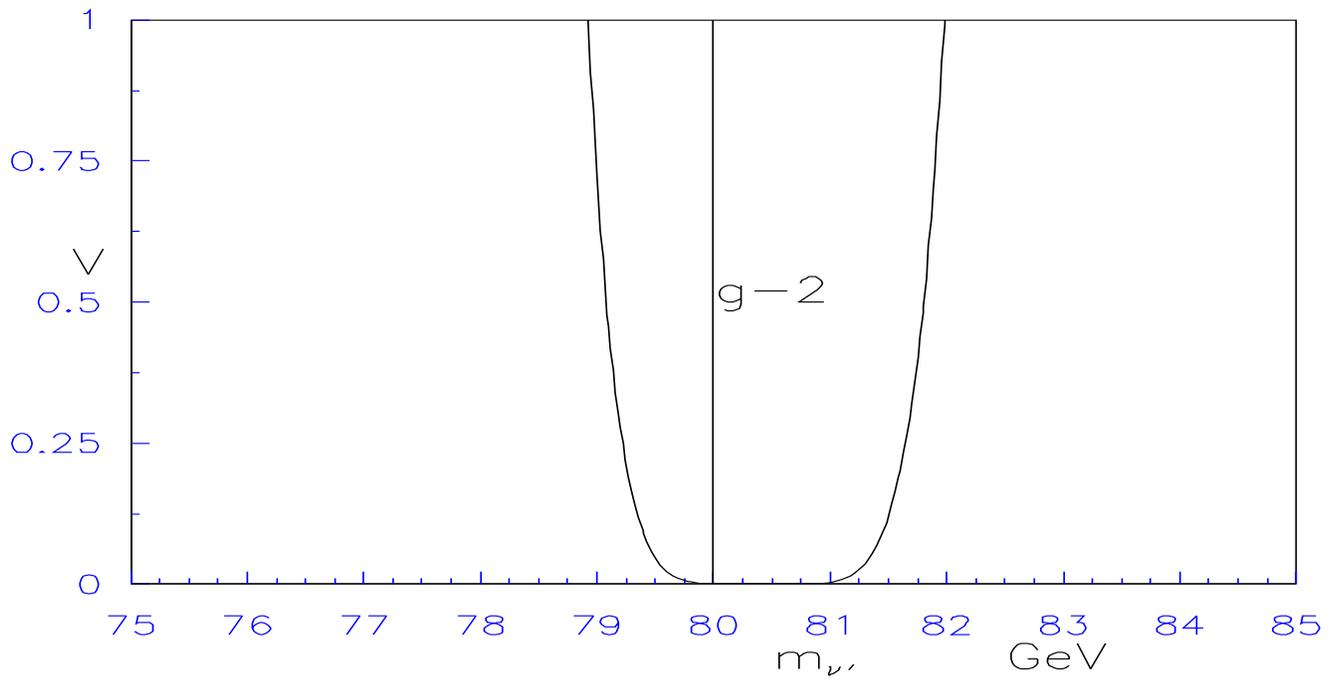}}
\caption{ Diagram of  $|V_{1\nu'} V_{2\nu'}|^2$, $|V_{1\nu'}
V_{3\nu'}|^2 $ and $|V_{3\nu'} V_{2\nu'}|^2$ to $m_{\nu'}$ }
\end{figure}

\begin{figure}
\vskip -4cm \epsfxsize=20cm \epsfysize=10cm \centerline{
\epsffile{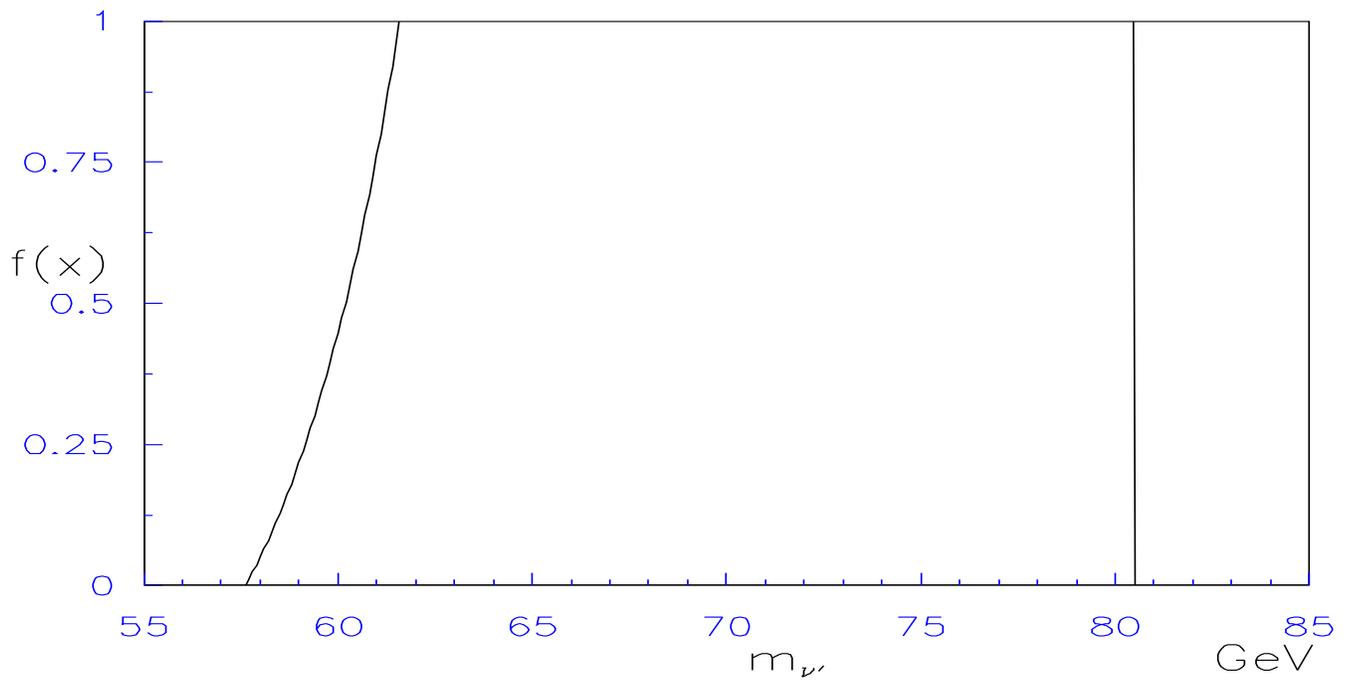}}
\caption{ Diagram of f(x) to $m_{\mu'}$ }
\end{figure}

\begin{figure}
\vskip -4cm \epsfxsize=20cm \epsfysize=10cm \centerline{
\epsffile{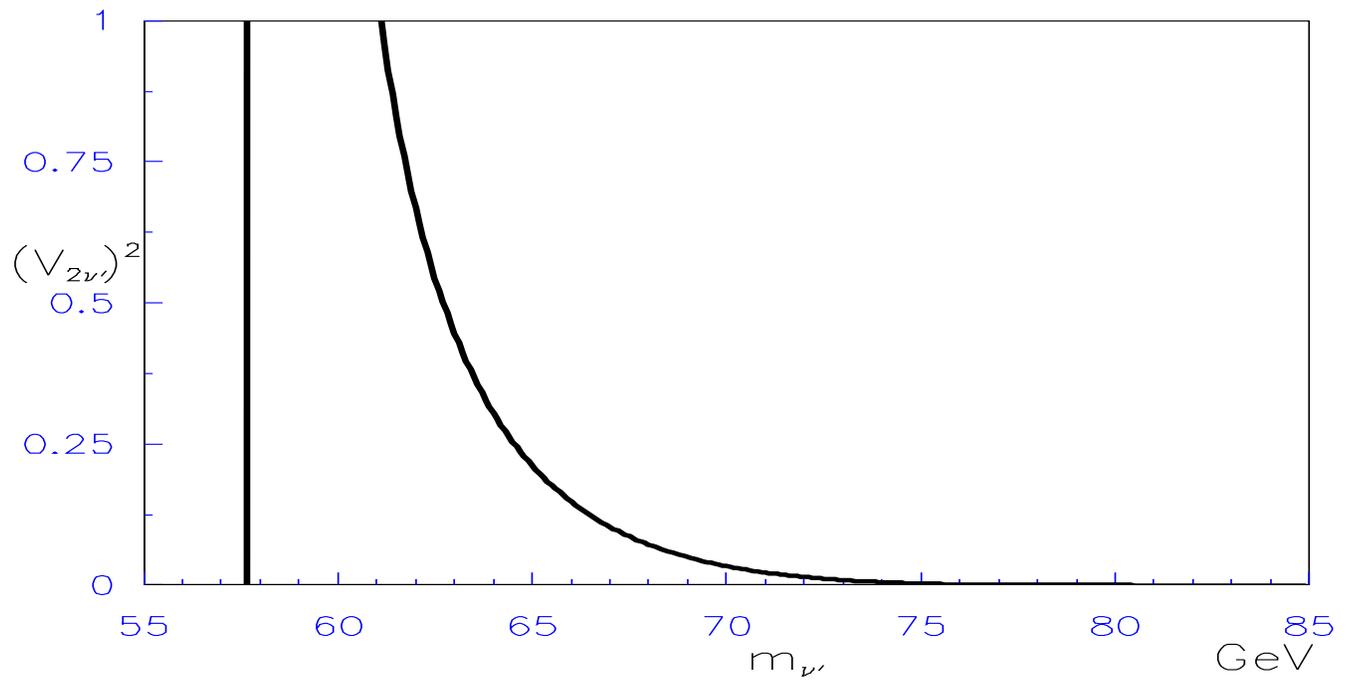}}
\caption{ Diagram of$V_{2\mu'}$ to $m_{\mu'}$ .}
\end{figure}

\end{document}